\documentclass[preprint,showpacs]{revtex4}
\usepackage{epsfig}
\usepackage{amsmath}

\begin{document}
\title{Critical properties of edge-cubic spin model on square lattice} 
\author{Tasrief  Surungan$^{1,2}$}
%\email{surungan@issp.u-tokyo.ac.jp}
\email{tasrief@unhas.ac.id}
\author{Naoki Kawashima$^{1}$}
\email{kawashima@issp.u-tokyo.ac.jp}
\author{Yutaka Okabe$^3$}
\email{okabe@phys.metro-u.ac.jp}
\affiliation{$^1$Institute for Solid State Physics, The University of Tokyo, Kashiwa, Chiba 277-8581, Japan \\
$^2$Department of Physics, Hasanuddin University, Makassar, South Sulawesi 90245, Indonesia\\
$^3$Department of Physics,
 Tokyo Metropolitan University, Hachioji, Tokyo 192-0397, Japan }

\date{\today}

\begin{abstract}
The edge-cubic spin model on square lattice
is studied via Monte Carlo simulation with cluster algorithm.
By cooling the system, we found two successive
 symmetry breakings, i.e., the breakdown of  $O_h$ into the group 
of $C_{3h}$ which then freezes into ground state configuration.
To characterize the existing phase transitions, we consider the magnetization
and the population number as order parameters. We observe that
the magnetization is good at probing the high temperature 
transition but fails in the analysis of the low temperature transition.
 In contrast the population number performs well in
  probing the low- and the high-$T$ transitions. We plot the
 temperature dependence of the moment and correlation ratios of
 the order parameters and obtain the high- and low-$T$ 
 transitions at $T_h = 0.602(1)$ and $T_l=0.5422(2)$
respectively, with the corresponding  exponents of correlation
length $\nu_h=1.50(1)$ and
$\nu_l=0.833(1)$. 
By using correlation ratio and size dependence of correlation
function  we estimate the decay exponent for the 
high-$T$ transition as $\eta_h=0.260(1)$.  For the low-$T$
transition, $\eta_l = 0.267(1)$ is extracted from the finite
size scaling of susceptibility.
The universality class of the low-$T$ critical point
is the same as the $3$-state Potts model.
\end{abstract}

\maketitle

\section{Introduction}
The presence of symmetry breaking in many areas of physics 
such as particle, atomic and condensed matter physics is indicative of
the importance of the phenomenon \cite{zinn}.
In general, the breakdown of symmetry is an onset of
a phase transition which separates phases with
different degrees of symmetry \cite{landau}. 
 A system is in high degree of symmetry 
at high temperature because it is able to explore 
all its configurational spaces. The decrease in temperature
will reduce thermal fluctuation and lead the system to stay
in some favorable states. This type of transition 
 with no co-existence phases, therefore no latent heat, 
is commonly called a continuous phase transition.

A system with initially large number of symmetry elements
 is more likely 
to experience sequential phase transitions. In fact,
various magnetic systems exhibit such behavior.
The clock spin model in two dimensions, for example, 
whose group symmetry $C_n$ experiences double Kosterlitz-Thouless
transitions for $n > 4$ \cite{jose}. 
In the presence of frustration which induces
chiral symmetry $Z_2$,  
 another phase transition occurs \cite{tasrief04}.

In this paper, we study the edge-cubic spin model on two
dimensions. The model is one of the discrete counterparts
 of the continuous-spin,
 the Heisenberg model, of symmetry group ($O_3$). In two 
dimension anisotropy is important as systems with discrete
symmetry can have a true long range order at finite temperature.
The octahedral symmetry group $O_h$
of the model, with 48 symmetry elements,
consists of some subgroups associated with
familiar discrete models, such as the inversion $Z_2$
of Ising model and the $C_{3h}$ of the chiral
3-state Potts model. Any finite ordered phase
 of the system is expected to be in one of 
 its subgroup symmetries.

While cubic symmetry in magnetic systems
is an old subject and appears whenever systems are 
on real cubic lattice \cite{aharony,kim}, 
cubic spin models have not been studied as much 
as other discrete models such as Ising, Potts and clock models. 
Previous works on cubic symmetries were mostly
carried out in the theoretical field approach
through the consideration of the $\phi^4$ Hamiltonian in which
$n$-component anisotropy fields break the continuous
$O(N)$ symmetry \cite{sznajd,calabrese02}.
It is well established that for three-dimensional case,
the cubic fixed point is stable if $n > n_c$, where
$n_c < 3$ according to more recent calculations \cite{calabrese02,carmona}.
The situation is different in two-dimensional case because 
the existence of cubic fixed point is still unclear.
Recent study by Calabrese {\it et al.}
could not unravel the speculation  that the Ising and
the cubic fixed points maybe coincide \cite{calabrese04}.

In trying to resolve the speculation of the
 existence of cubic fixed point in two
dimensions, it is of importance to directly 
probe the spin models with cubic symmetry.
With simple normalized-vector spin on  cube, we can 
have three models  i.e, the face-cubic (6 states),
the corner-cubic (8 states) and the edge-cubic spin (12 states).
Very recently,  Yasuda {\it et al.} \cite{yasuda} considered
a ferromagnetic face-cubic spin model
 and then found that the model undergoes a single 
phase transition; they discussed  the  universality 
class of this  model in comparison with 4-state Potts model. 
The corner-cubic spin model is considered
as a  trivial model of  decoupled 
three independent Ising models. Studying the corner-cubic
model can not be expected
to address the existence of cubic fixed
point. However, by weighting the spin orientation,
the corner-cubic spin model can transform into
a general Ashkin-Teller model \cite{ashkin}; to this respect the 
model is no longer trivial.

Probing the edge-cubic spin model deserves for its own right.
 Firstly, it is interesting to know
the symmetry breaking of the $O_h$ in that model, and
also to address the existence of cubic universality class.
Since the degree of $O_h$ is high, the edge-cubic model
may experience sequential phase transitions.

The remaining part of the paper is organized
as follows: Section II describes the model and the method. 
The result  is  discussed in Section III. Section IV is devoted
to the concluding remarks.  

\section{Model and Simulation Method}

The edge-cubic spin model is one of the discrete counterparts of the Heisenberg
model. Spins can point to any of the 12 middle points of the edges of 
a cube. 
An edge of the cube is 2 unit long, 
and its center of mass $O(0,0,0)$ is set 
  as the origin of the normalized-vector spin.
Here we study the ferromagnetic case on square lattice with
periodic boundary condition. 
The Hamiltonian of the model is expressed as
\begin{equation}
H = - J \sum_{\langle ij \rangle}\vec s_i \cdot \vec s_j 
\end{equation}
where $\vec s_i$ is a spin on site $i$-th,  $J > 0$. Summation is performed over all
the nearest-neighbor pairs of spins. 
In the ground state configuration, i.e., when all spins having a common
orientation, the energy will be $-2JN$  with $N$ is the number of spins.

%--------------------------------fig01--------------------------------------
\begin{figure}[t]
\includegraphics[width=1.0\linewidth]{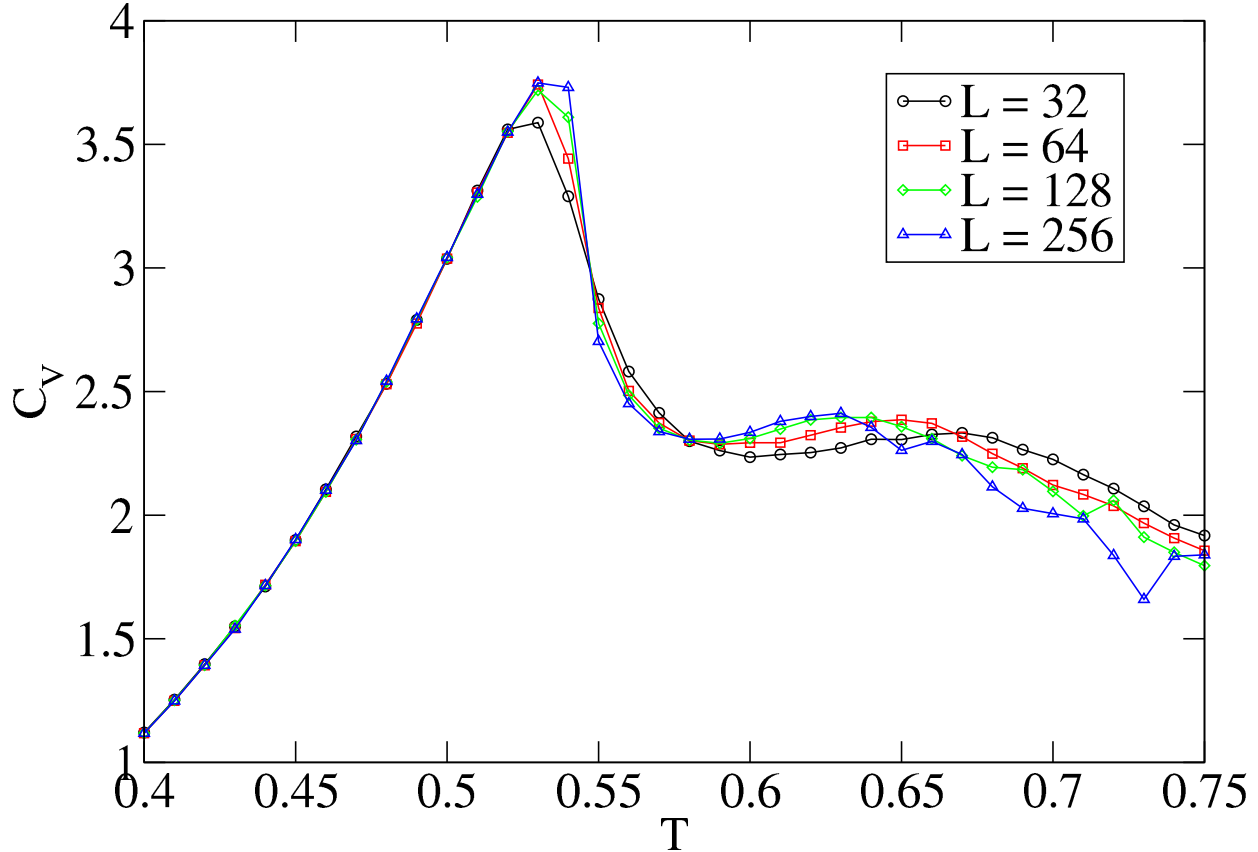}
\vspace{0.1cm}
\caption{The temperature dependence of the specific heat for
various system sizes. As shown, there exist a clear peak and a hump 
respectively at high and at low  temperatures. 
 The peak and the hump signify that system possesses 
sequential phase transitions. The error bar, in average, is in the order of
symbol size, except data for  $L=256$ at $T>0.65$, where error is larger 
than the symbol size.}
\label{SPHT}
\vspace{0.6cm}
\end{figure}
%------------------------------------------------------------------------------

We use the canonical Monte Carlo (MC) method with  single
 cluster spin updates due to Wolff \cite{wolff} and 
adopt Wolff's idea of embedded scheme in constructing a cluster
for the edge-cubic spins. This is done by projecting the spins
into a randomly generated plane so that the spins 
are divided into two groups (Ising-like spins).  
The embedded scheme is essential in carrying out cluster algorithm for 
such spins as cubic and  planar spins.

After the projection, the usual steps of the cluster algorithm are 
 performed \cite{kasteleyn},
i.e., by connecting bonds from the randomly
chosen spin to its nearest neighbors of similar group, 
with suitable probability. 
This procedure is repeated for neighbors of sites 
connected to chosen spin until no more spins to include.
One Monte Carlo step (MCS) is defined as visiting once the whole 
spins randomly and perform cluster spin update in each visit.
Note that a spin may be updated many times, in average,
during one step, in particular near the critical point.

Measurement is performed after enough equilibration MCS's ($10^4$ MCS's). 
Each data point is obtained from the average over several parallel
runs, each run is of $4 \times 10^4$ MCS's. 
To evaluate the statistical error each run is treated as a single measurement.
For the accuracy in the estimate of critical exponents and 
temperatures,  data are collected  upto
more than 100 measurements for each system size.

\section{Results and Discussion}

\subsection{Specific heat and magnetization}

The first step in search  for any possible phase transition
is to measure the specific heat of the system defined as follows

\begin{equation}
C_v(T) = \frac{1}{Nk_BT^2}(\langle E^2\rangle - \langle E\rangle^2)
\end{equation}
where $E$ is the energy in unit of $J$  
 while $\langle \cdots \rangle$ represents
the ensemble average of the corresponding quantity.
All temperatures are expressed in unit of $J/k_B$ where
$k_B$ is the Boltzmann constant.

%--------------------------------fig2--------------------------------------
\begin{figure}
\includegraphics[width=1.0\linewidth]{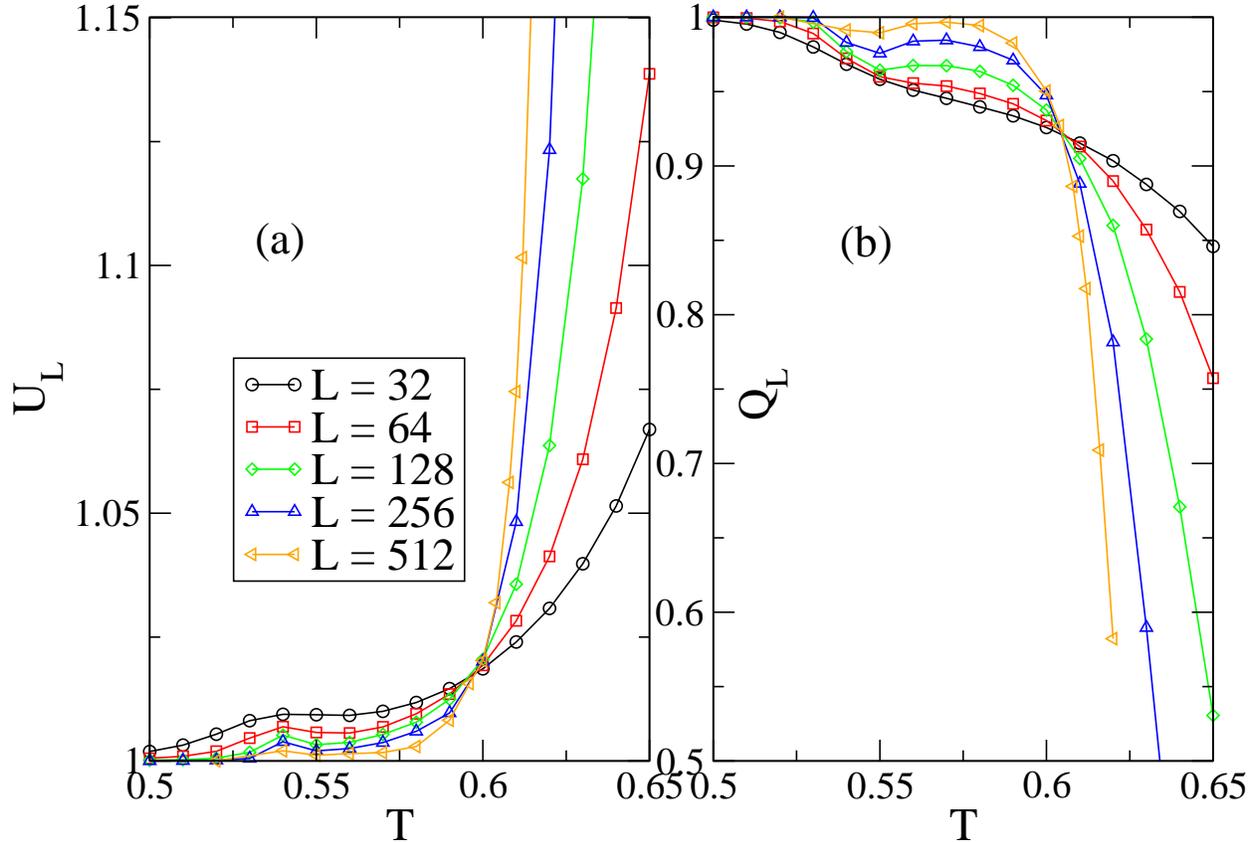}
\vspace{0.1cm}
\caption{Temperature dependence of (a) moment ratio and (b)
correlation ratio for several system
sizes. The crossing points indicates a phase transition between
the disordered  and the intermediate phase. A cusp in the moment ratio and valley-like in the correlation ratio suggest
 another phase transition. Error bar is in the order smaller than the symbol size.}
\label{RATIO_m}
\vspace{0.5cm}
\end{figure}
%------------------------------------------------------------------------------

As shown by the specific heat plot in Fig. \ref{SPHT},
 there exist a peak at lower temperature
and a hump at higher temperature. Although the  peak and the hump 
are more directly related to energy fluctuation,
they may signify the existence of sequential phase transitions.
In what follows, more quantitative analysis is performed 
through the evaluation of the order parameters.

The critical properties of the system are quantified by the
critical temperatures and exponents extracted from the finite size 
scaling (FSS) of the order parameters, in particular from their
 moment and correlation ratios.
As the probed system is ferromagnetic we consider magnetization
$M = |\sum \vec s_i|$ as the order parameter. By defining 
$M^k$ as the $k$-th order moment of magnetization and
$g(R) = \sum \vec s(r) \cdot \vec s(r+R)$ as correlation function,
the moment and correlation ratios are respectively written as 
follows
\begin{eqnarray} 
U_L &=&\frac{\langle M^4\rangle}{\langle M^2\rangle^2}\\
Q_L &=&\frac{\langle g(L/2)\rangle}{\langle g(L/4)\rangle}\\
\nonumber
\end{eqnarray} 
Precisely, the distance $R$ for the correlation function $g(R)$
is a vector quantity, here we take the simple form and choose 
convenient distances $L/2$ and $L/4$, both in $x$- and $y$-directions.

%--------------------------------Fig03--------------------------------------
\begin{figure}
\includegraphics[width=1.0\linewidth]{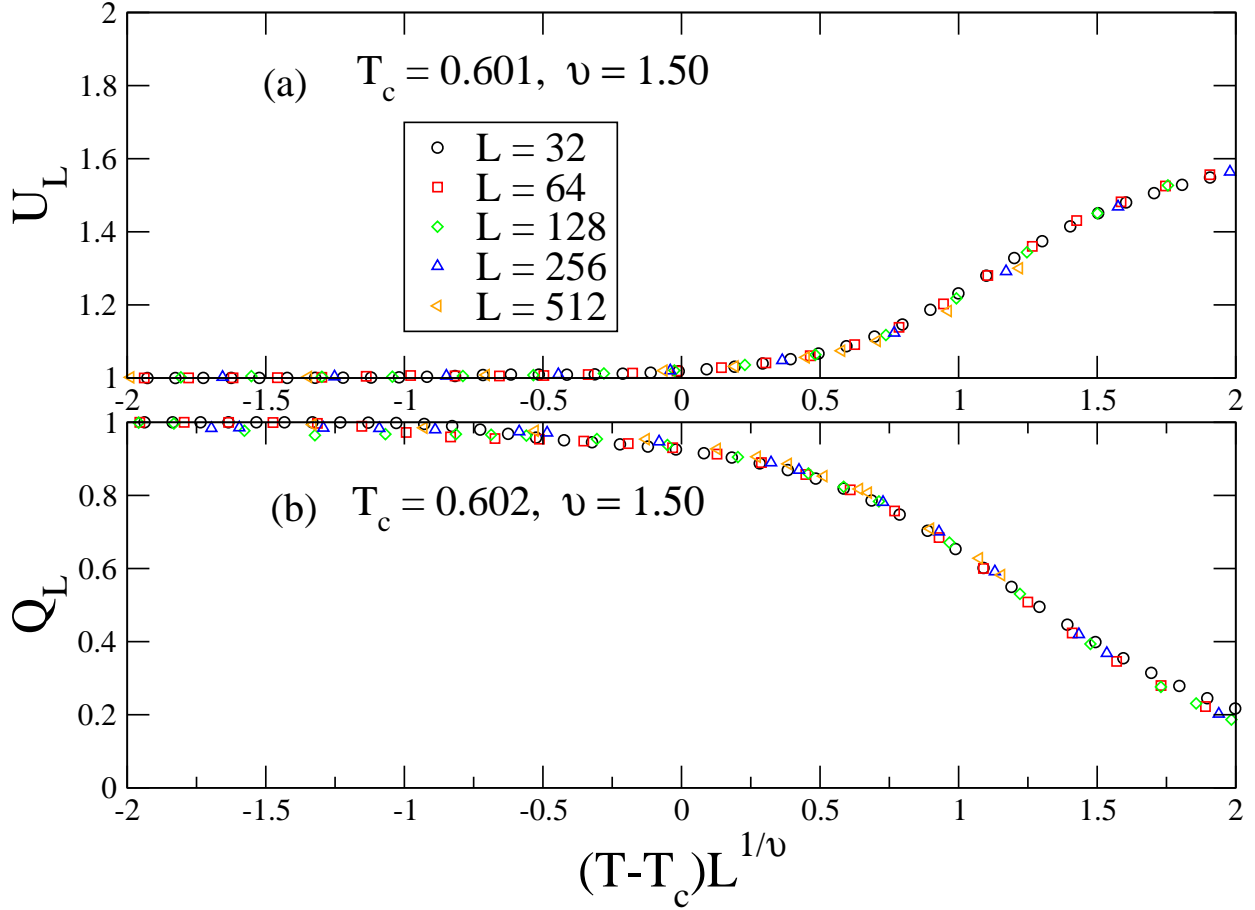}
\vspace{0.1cm}
\caption{The FSS plot of (a) moment ratio, (b) correlation ratio. The estimates of critical temperature and the exponent of correlation
length $\nu$ are obtained.}
\label{BINDER_m_s}
\vspace{0.5cm}
\end{figure}
%------------------------------------------------------------------------------

%--------------------------------Fig04--------------------------------------
\begin{figure}
\begin{center}
\includegraphics[width=0.9\linewidth]{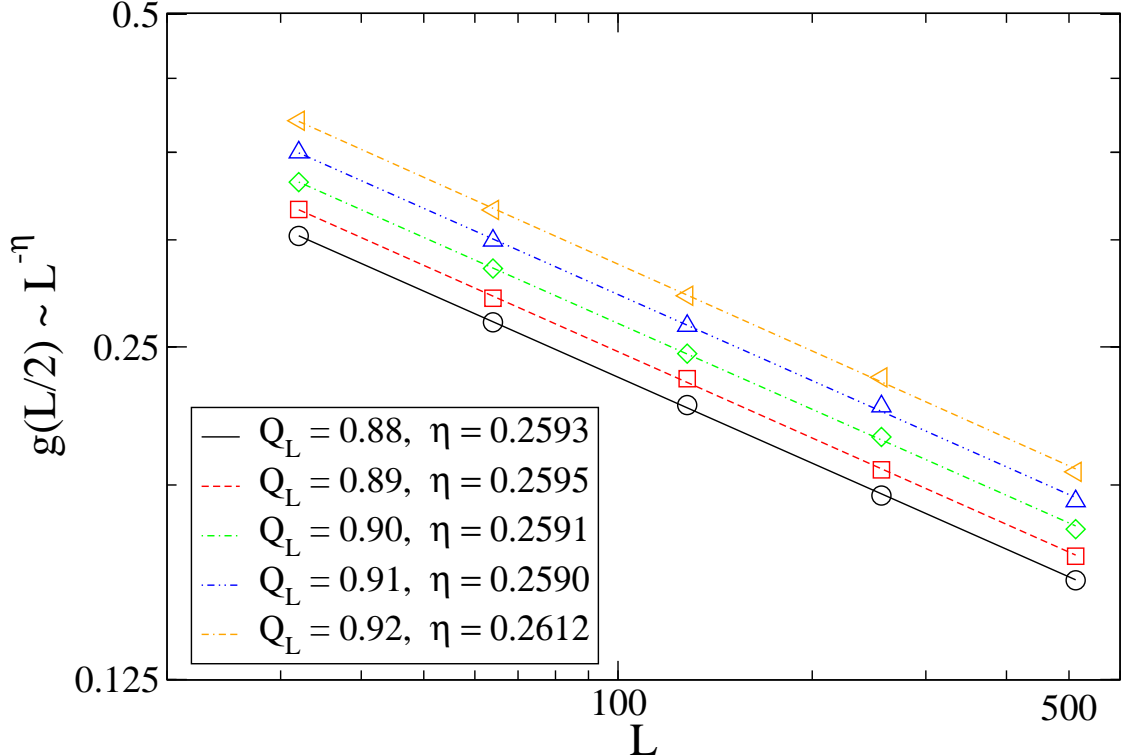}
\vspace{0.5cm}
\caption{Double logarithmic plot of $g(L/2)$ vs $L$ for
several values of correlation ratio $Q_L$'s.
The best estimate of $\eta = 0.260(1)$ is the slope
 of the best-fitted line of $Q_L=0.91$ associated with critical temperature.
}
 \label{eta_ord1}
\end{center}
\end{figure}
%------------------------------------------------------------------------------

More accurate estimate of parameters of phase transition
is obtained from the temperature dependence
of $U_L$ and $Q_L$. At very low temperature where system is approaching 
the ground state, both moment and correlation ratio are trivial.
 Due to the absence of fluctuation,
 the distribution of $M$ is a delta-like function,
  giving moment ratio equals to unity.
Correlation ratio also goes to unity as correlation
function for small and large distance is the same 
due to  highly correlated states. 
In excited states, the moment and the correlation
ratios are not trivial, they depend on temperature.
The plot of moment ratio for various system sizes, shown
in Fig. \ref{RATIO_m}(a), exhibits a clear crossing point which indicates
a phase transition. At low temperature side, there exists
 a cusp which may 
corresponds to another phase transition. 
A possibility that system has additional phase transition at low temperature,
apart from an obvious one at high temperature, is also
signified by the plot of correlation
ratio shown in Fig. \ref{RATIO_m}(b).

We show the FSS plot of moment and correlation ratios in 
Fig. \ref{BINDER_m_s}; we estimate critical temperature and exponents from
both ratios which give consistent results,  with only
difference smaller than estimated statistical error.
 The estimate of $T_c$ obtained from moment 
ratio is $T_c = 0.601(1)$, slightly smaller than $T_c = 0.602(1)$
 from the correlation  ratio. 
The number in bracket is the uncertainty in the last digit. 
In general, moment ratio has larger correction
to scaling than the correlation ratio \cite{tomita02a}, which happens
to be the case here.
However, if the variables of the two correlation functions
are not local  quantity, in the sense 
they depend on another quantity, then
the correlation ratio may have larger correction to scaling.
Our estimate of $T_c$ is based on result obtained
from the correlation ratio. The estimates of the decay 
exponent of the correlation
length both give the same results, i.e.,  $\nu_h = 1.50(1)$.
 The subscript is used for the reminder
that we are dealing with the high-$T$ transition.

In addition to the exponent $\nu$, it is possible to extract
the decay exponent $\eta$ of the correlation function
 from the correlation ratio.
This is done by firstly looking at the constant value of correlation ratio $Q_L$
for different sizes and then find the corresponding correlation function $g(L/2)$.
The correlation function is in power-law dependence on 
the system size, $g(L/2) \sim L^{-\eta}$ \cite{tomita02a}.
Therefore, if we plot $g(L/2)$ versus $L$ for various $Q_L$'s
in logarithmic scale, as in Fig.~\ref{eta_ord1},
the value of $\eta$ will correspond to the gradient of
the best-fitted line for each constant of correlation ratio.
There are several lines plotted in Fig. \ref{eta_ord1}. Since
the critical temperature is associated with the value of $Q_L \sim 0.91$
(Fig. \ref{RATIO_m}(b)), we assign $\eta = 0.260(1)$ as the best estimate. 

%--------------------------------fig05--------------------------------------
\vspace{1cm}
\begin{figure}
\includegraphics[width=0.5\linewidth,height=0.5\linewidth]{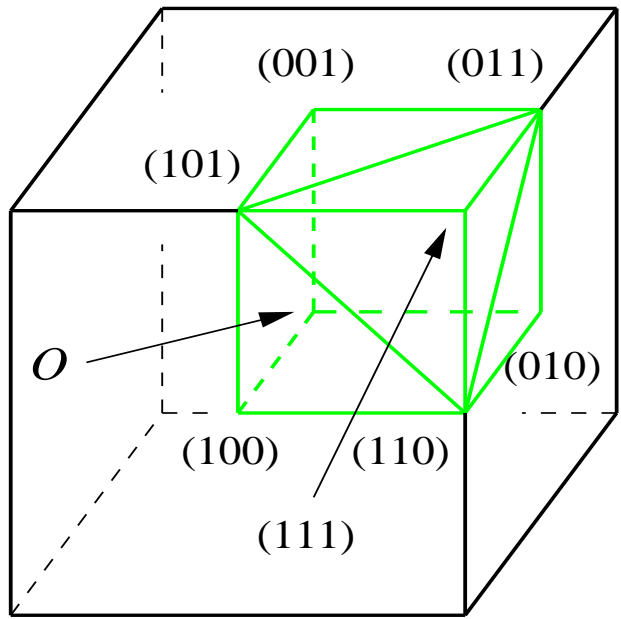}
\caption{The cube and 1/8 cube (green). Three outer
sides of the small cube  viewable from corner (111) are 
the surfaces penetrated by  series of magnetization vectors. 
}\label{CUBE}
\end{figure}
%------------------------------------------------------------------------------

Although there is an indication of low temperature transition, 
at this stage we do not estimate its critical quantities 
 due to the absence of a crossing point.
We need to formulate a more suitable order parameter able
to  distinguish the intermediate and the low temperature ordered phase.
In the next section, we present the snapshot series of total
 magnetization and discuss the order parameter of characterizing
low temperature transition.

\subsection{Snapshot of spin configuration and population number
 order parameter}
%--------------------------------fig06--------------------------------------
\vspace{1cm}
\begin{figure}[t]
\includegraphics[width=0.9\linewidth,height=0.9\linewidth]{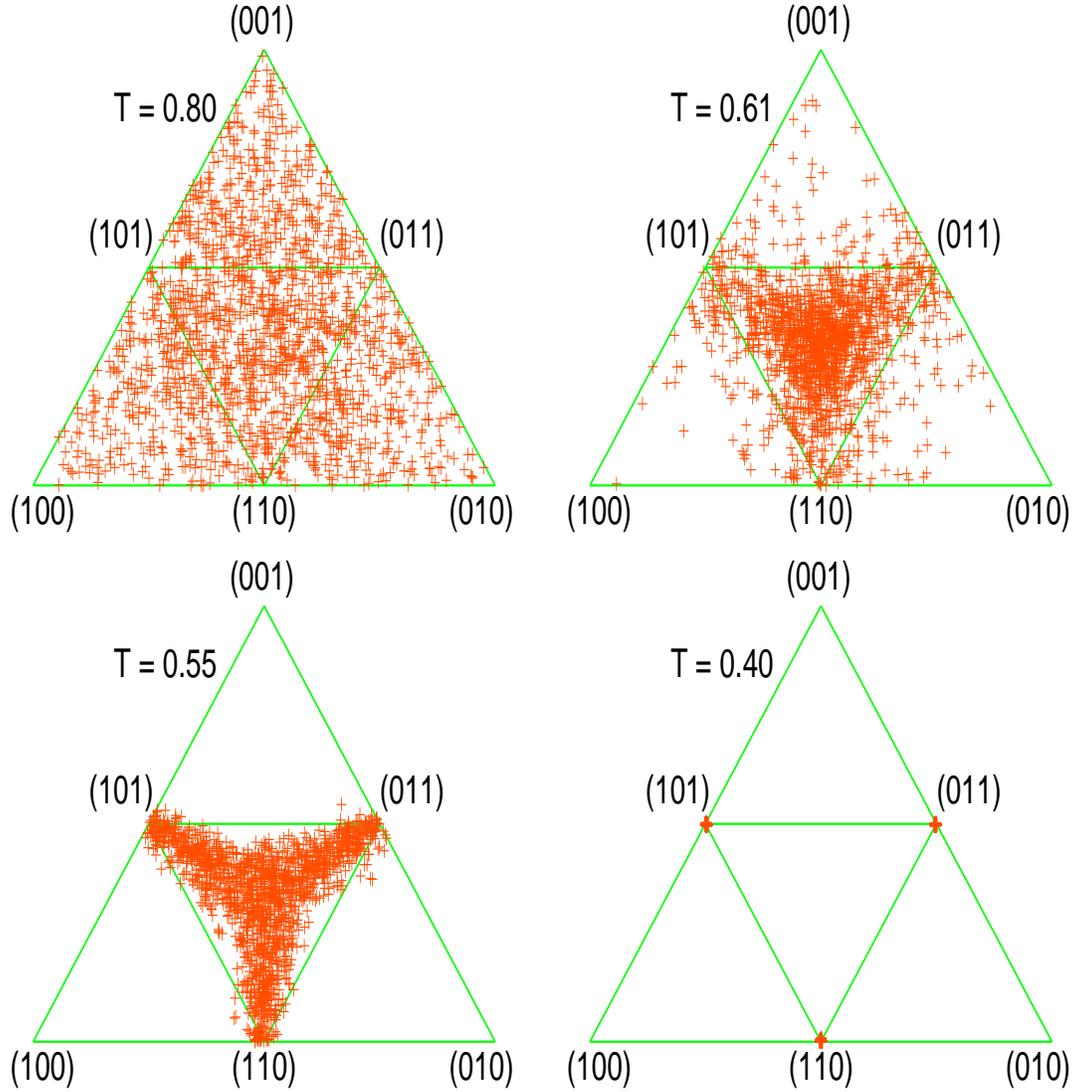}
\caption{Plot of the series of snapshots of total magnetization.
 The orientation is viewed from the corner point (111) of the
cube. Each symbol represents the total magnetization 
of a snapshot spin configuration.}\label{SNAP}
\vspace{0.5cm}
\end{figure}
%------------------------------------------------------------------------------

The total vector magnetization is computed 
for every snapshot spin configuration.
A snapshot magnetization is represented by a dot
that is the intersection of the line parallel
to the magnetization and the cube surface.
Thus we obtain dots as many as the number 
of MCS's, and we view this dots
from the (111) direction as shown in 
Fig. \ref{CUBE}.

%--------------------------------fig07--------------------------------------
\vspace{1cm}
\begin{figure}
\includegraphics[width=0.9\linewidth,height=0.7\linewidth]{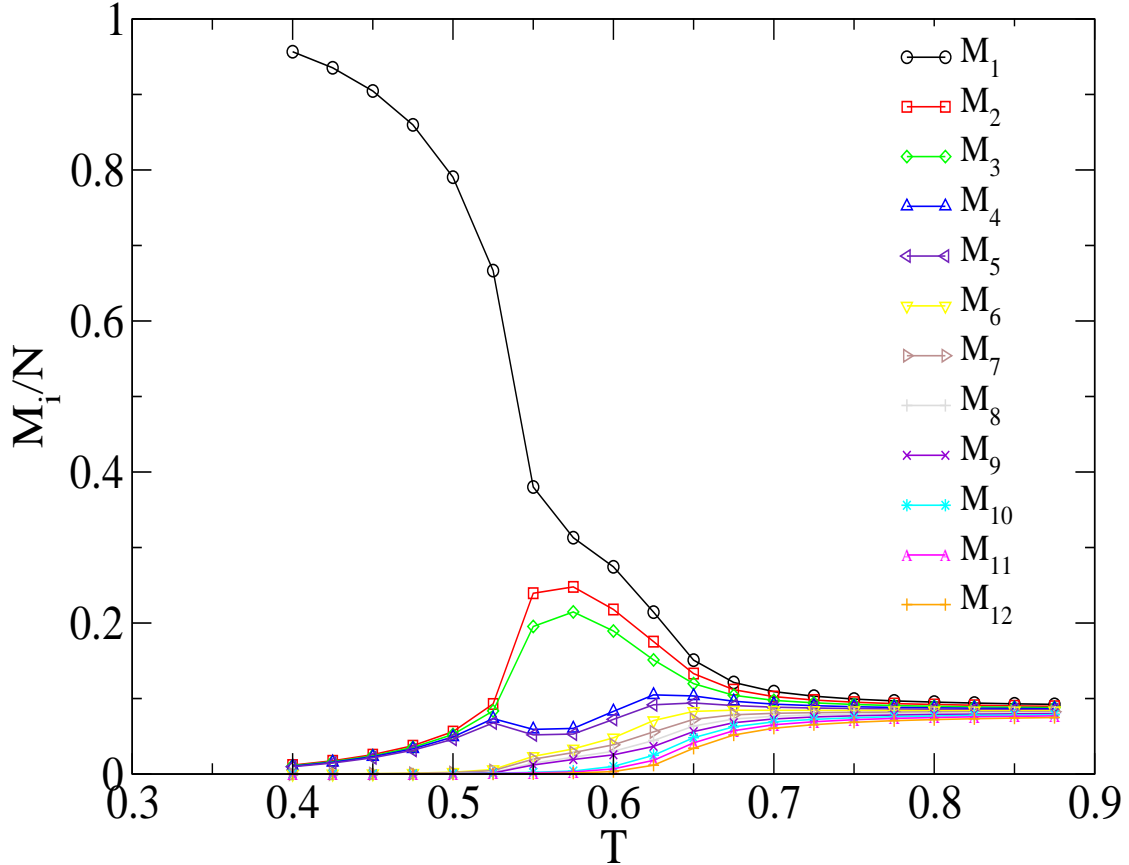}
\caption{Plot of temperature dependence of population number, $M_i$
i.e., the thermal average of number of spins pointing to $i$-th direction.
At each MCS, 12 possible directions are sorted and we assign the most
populated as $M_1$, the second largest as $M_2$, and so on. Here
the linear size of the lattice is $L=128$.}\label{Ni}
\end{figure}
%------------------------------------------------------------------------------

For simplicity, we make suitable mirror projection of the 
 total magnetization so that its orientation 
 is in the region of the three outer sides of the 1/8 cube.
We further project the surface of the 1/8 cube on to 
a triangle as shown in Fig. \ref{SNAP}.
The inner triangle is associated with  the plane made by 
three edge-points ((101), (110) and (011)) in Fig. \ref{CUBE}.
Each pair of these points together
with a  middle point of sides viewable from
the corner point (111) construct three other 
 outer triangles.

The phase of the system is related to spin configuration.
 At high temperatures, due to large thermal fluctuation, each 
spin is relatively free to point to any direction, therefore
no common orientation of the
total magnetization.  As a result, the  snapshot point
will occupy the whole area of four triangles. 
This is indicated by the Fig. \ref{SNAP} with $T=0.80$.

As we reduce the temperature,
the thermal fluctuation starts being overcome by
the magnetic interaction. The snapshot points start
being around the middle area of the triangle (associated with
$T = 0.61$). At this state, three neighboring spin-orientations
near a corner of the cube become more favorable.
At temperature $T=0.55$ 
system is in intermediate phase where almost all snapshots
are inside the area of the inner triangle.
Three neighboring orientations around a particular
corner of the cube are chosen; the octahedral
symmetry $O_h$ is completely broken.

The symmetry group associated with the intermediate 
phase is the point group $C_{3h}$, realized for example by 
the 3-state Potts model.
 As the temperature is further reduced, this
symmetry breaks down into a ground state
 with all spins pointing to the same direction,
 shown by the figure with $T=0.40$.
 Therefore, from the symmetry group point of view, it is natural
to expect that the low temperature phase transition
is in the same universality class as the 3-state Potts model.

%--------------------------------Fig08--------------------------------------
\begin{figure}
\begin{center}
\includegraphics[width=1.0\linewidth]{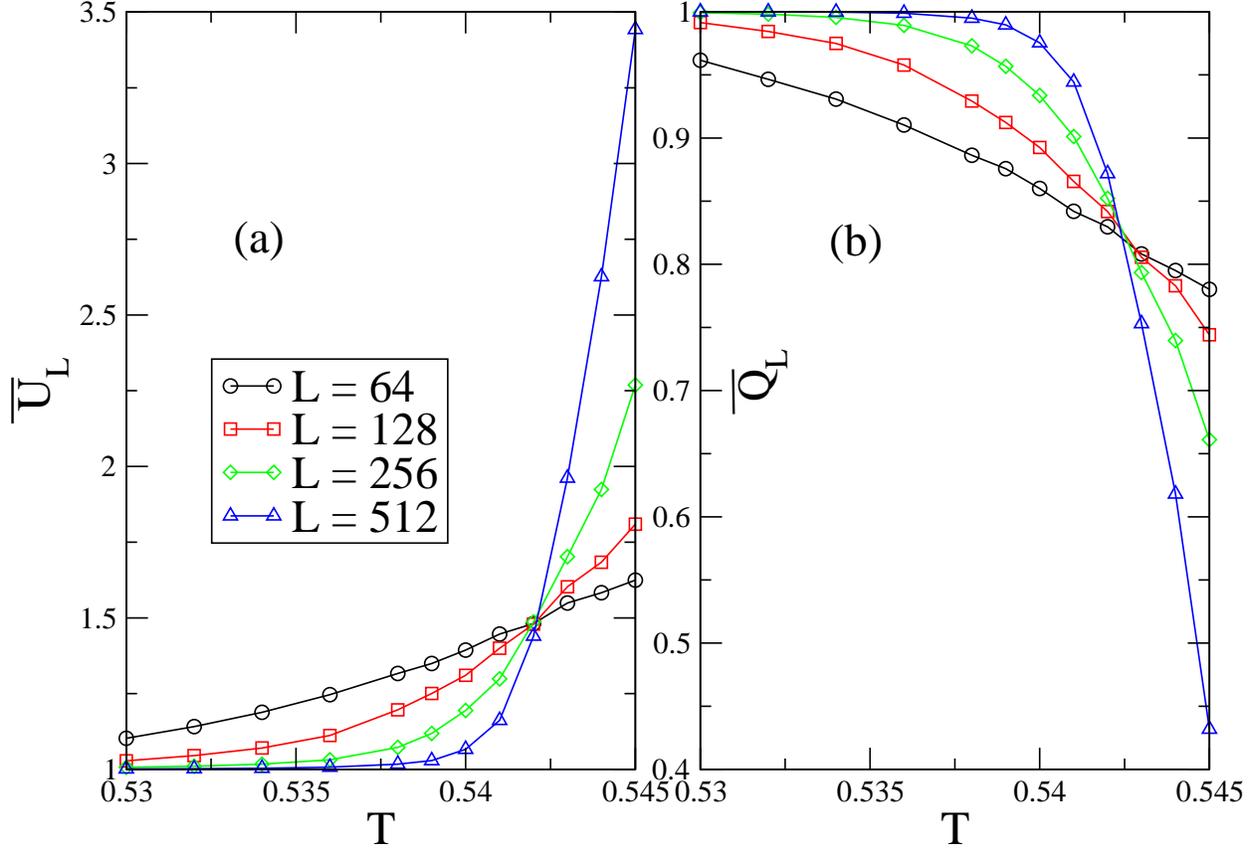}
\vspace{0.5cm}
\caption{Temperature dependence of (a) moment ratio and (b)
correlation ratio for the occupation number order parameter defined in Eq. 
(\ref{ordp2}). Error bar is in the order of symbol size.}
\label{ratio2}
\end{center}
\end{figure}
%------------------------------------------------------------------------------
 
Based on the snapshot of magnetization, we define an order
parameter related to  population numbers. It is formulated
from the fact that at each microscopic state, spins will
be pointing to 12 possible orientations.
The difference between maximum population among the 12 
orientations and the second largest is assigned as the order 
parameter which is written as follows
\begin{equation}\label{ordp2}
\bar M = M_1 - M_2 
\end{equation}
where $M_1$ and $M_2$ are the largest
and the second largest population
numbers, respectively.

At ground state the value of this order parameter 
will be just $N$ because all $N$ spins are in a common 
alignment.  In contrast, at high temperature, the value of
 the order parameter is very small and vanishing in the thermodynamic limit,
as the 12 possible orientations are occupied  by 
approximately equal number of spins.  Figure \ref{Ni}
shows the temperature dependence of the 12 population numbers $M_i$.
One could choose another quantity as an order parameter, but Eq. \ref{ordp2} 
is simple and straightforward.

The  breakdown of symmetries experienced by the system
can also be detected from temperature dependence of $M_i$.
 At high temperature side ($T > 0.8$),
12 lines are approximately parallel. 
There is a clear split of lines  at temperature around $T=0.61$, 
 where 3 lines go up whereas the others go down. 
This reminds us of Fig. \ref{SNAP} with $T=0.61$, where
three neighboring spins around a cube corner start being
favorable. At lower temperatures ($T < 0.55$), 
the upper three lines separate into two groups, where
one continuously goes up while the other two are
vanishing. This exhibits the breakdown of $C_{3h}$.

%--------------------------------Fig09--------------------------------------
\begin{figure}
\begin{center}
\includegraphics[width=1.0\linewidth]{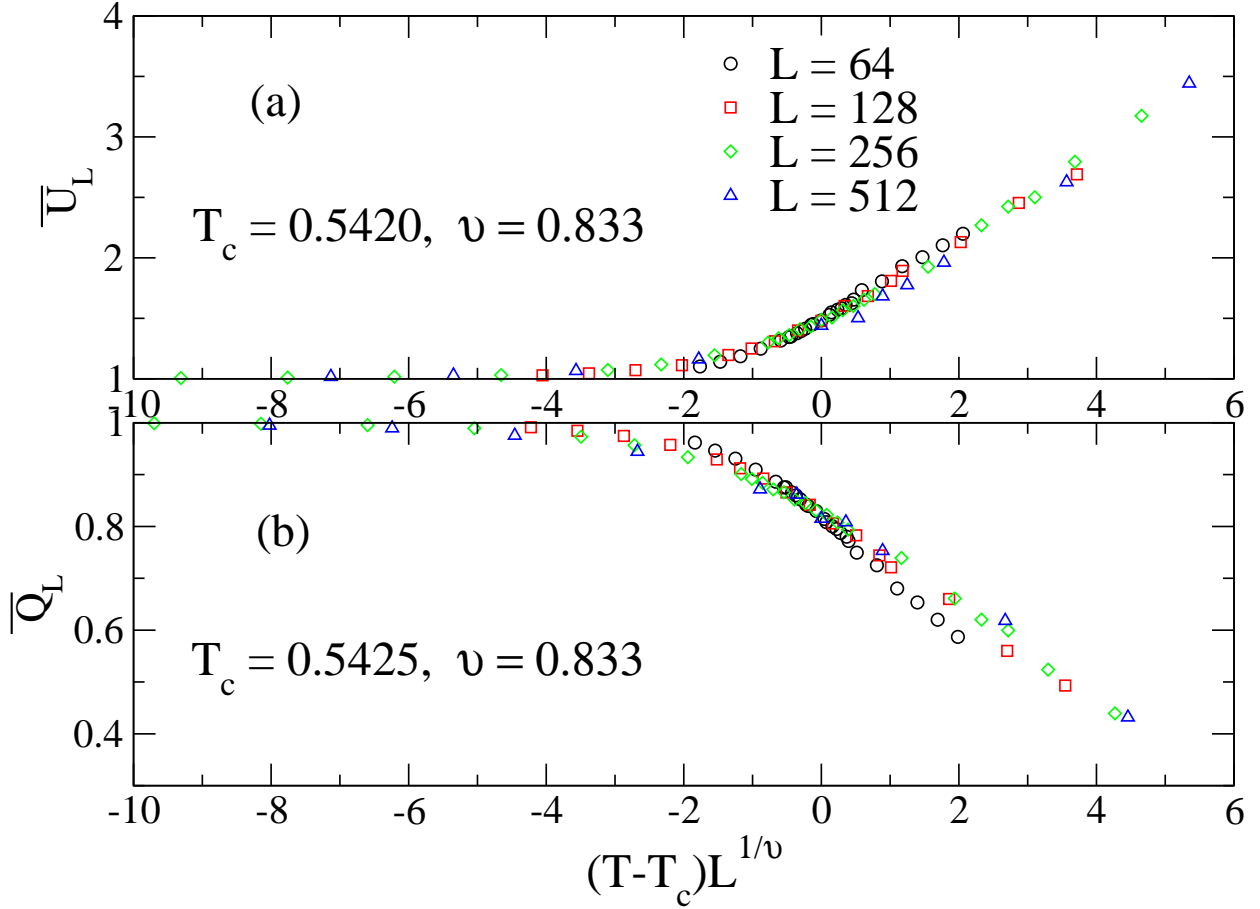}
\vspace{0.5cm}
\caption{Finite size scaling plot of  (a) moment ratio and (b)
of Fig. \ref{ratio2}.}
\label{fss_L}
\end{center}
\end{figure}
%------------------------------------------------------------------------------

In this part, we make use the moment and the correlation
 ratio of the newly introduced order parameter for
the analysis of the low temperature critical behavior,
 expressed as
\begin{eqnarray} 
\bar U_L &=&\frac{\langle \bar M^4\rangle}{\langle \bar M^2\rangle^2}\\
\bar Q_L &=&\frac{\langle \bar g(L/2)\rangle}{\langle \bar g(L/4)\rangle}\\
\nonumber
\end{eqnarray} 
Here, the correlation function $\bar g(R) = 
\sum  \bar M(R+r) \bar M(r) $, where $\bar M(R)$
 is defined as follows
\begin{equation}
\bar M(R) = M_1(R) - M_2(R)
\end{equation}
where $M_k(R)$ is 0 or 1. 
When the spin at $R$, $\vec s(R)$, is in the $k$-th 
populated direction
$M_k(R)=1$, otherwise it is 0.
This means the direction 1 (or 2) can be different
for different Monte Carlo steps. With this
definition, the correlation function relates points with
functional variable of occupation number. 
The plot of these ratios for various system sizes, given in Fig. \ref{ratio2},
 has shown a clear 
 crossing point separating between the intermediate and the low temperature
order phases. The FSS plot
of the moment and correlation ratios, shown in Fig. \ref{fss_L}, gives
the estimates of $T_c$ and the exponent $\nu$.
While the quality of the plot is not as good as
that of the susceptibility discussed below,
we estimate  $T_c = 0.5422(2)$ and $\nu=0.833(1)$ based on moment ratio.
This exponent is in a good agreement with 
the result of 3-state Potts model \cite{wu}.

\begin{figure}
\begin{center}
\includegraphics[width=0.9\linewidth,height=1.1\linewidth]{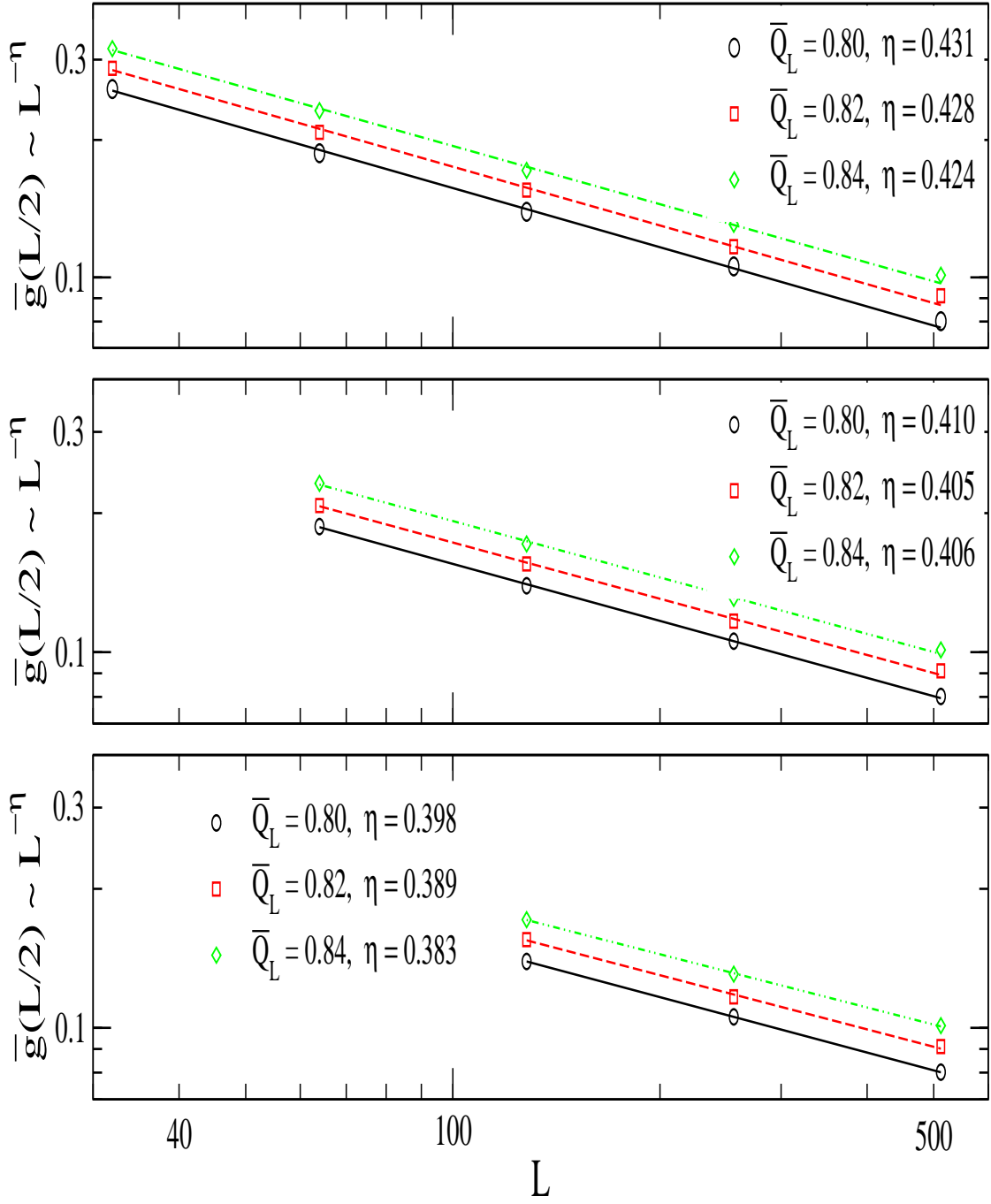}
\vspace{0.5cm}
\caption{Double logarithmic plot of $\bar g(L/2)$ vs $L$. The
gradient of the fitted line associated with $\bar Q_L=0.82$ is the
 estimate of $\eta$. By excluding the smaller system sizes, 
the value of estimated $\eta$ systematically increases,
which indicates large correction to scaling.}
\label{eta_ord2}
\vspace{0.4cm}
\end{center}
\end{figure}

The decay exponent $\eta$ for the
correlation function of  $\bar M$ can also
be extracted in the same way for $M$. 
After determining a fixed value of correlation ratio,
 we search for  $\bar g(L/2)$ of
each system size and plot against $L$
in logarithmic scale. The  gradient of the
best-fitted line associated with critical value of 
correlation ratio is the  estimate of $\eta$.
However, due to large  correction to scaling of the correlation
ratio, the estimates of $\eta$
is off from the 3-state Potts model.
By excluding the small system sizes, as indicated
in Fig. \ref{eta_ord2},
the estimated value of $\eta$ systematically
declines. We believe that the
value of the 3-state Potts model is approached
for  larger system sizes.

\begin{figure}
\begin{center}
\includegraphics[width=1.0\linewidth]{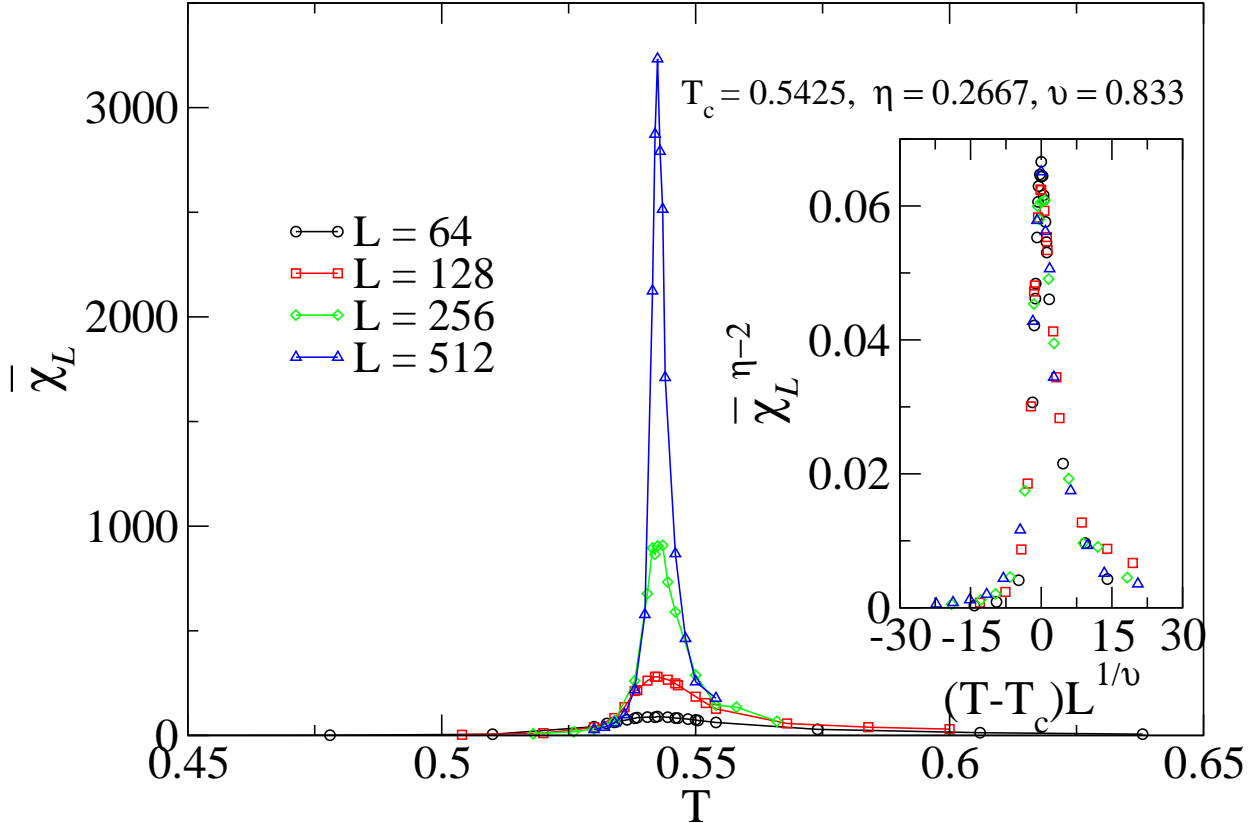}
\vspace{0.5cm}
\caption{Temperature dependence of susceptibility and
its FSS (inset). The exponents  
$\eta $ and $\nu$ are very consistent with
 that of 3-state Potts model.}
\label{chi_ordp2}
\end{center}
\end{figure}

In order to obtain a better estimate of  $\eta$,
we perform another approach, namely by using
the FSS of susceptibility $\chi_L$, which is written as follows
\begin{equation}
\bar \chi_L = L^{2-\eta} \tilde{\bar\chi}_L \left((T-T_c) 
L^{1/\nu} \right),
\end{equation}
where
$
\bar\chi_L = \left(\langle \bar M^2\rangle 
- \langle \bar M \rangle^2 \right)/L^2
$.
%\end{equation}
The temperature dependence of $\bar \chi_L$ is given in Fig. \ref{chi_ordp2}
with the inset is its FSS plot. The exponent
$\nu = 0.833(1)$ and $\eta = 0.267(1)$ belong to
 3-state Potts model.

%--------------------------------Fig10--------------------------------------
%------------------------------------------------------------------------------

\begin{figure}
\begin{center}
\includegraphics[width=0.9\linewidth]{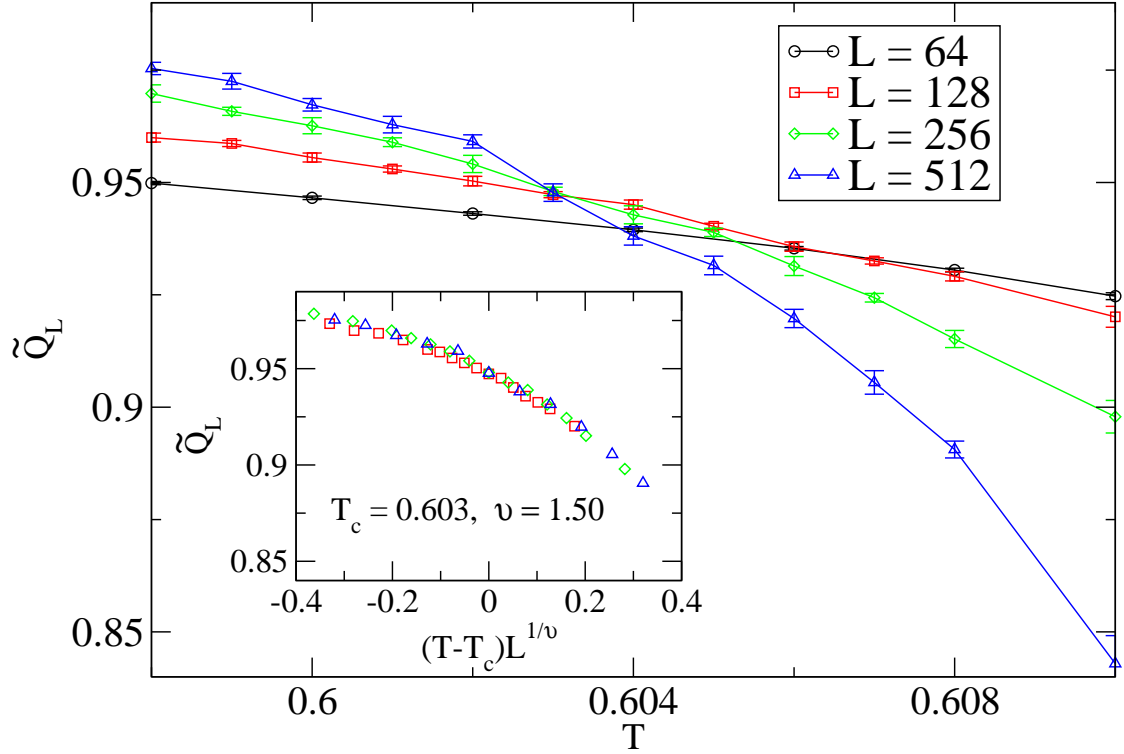}
\vspace{0.5cm}
\caption{Correlation ratio of the order parameter $\bar M = M_1-M_4$
vs temperature.  There is a clear crossing point for
curves of larger system sizes, $L$ = 128, 256, and 512. The system size
$L =$ 64 seems to have quite large correction to scaling.
The inset is the FSS for the 3 large system sizes.}
\label{corr_ord3}
\vspace{0.6cm}
\end{center}
\end{figure}

Next, we address the possibility of characterizing
the high temperature transition by using
the order parameter based on population number.
Obviously, the order parameter introduced in Eq. (\ref{ordp2})
is only appropriate  for low temperature,
not for high temperature transition.
This is due to the fact that at high and intermediate temperatures,
 $M_1$ and $M_2$ have approximately similar value, especially
for larger system sizes. 
In the intermediate phase, the three neighboring spin-orientations
are favorable; thus for probing high temperature transition,
it is appropriate to subtract $M_4$, 
the fourth largest population number, instead of $M_2$,
from $M_1$, and obtain $\tilde M = M_1 - M_4$, analogous to 
Eq. (\ref{ordp2}). The temperature dependence
of the correlation ratio and its FSS is shown in Fig. \ref{corr_ord3}.
 The plot of temperature dependence for various 
system sizes gives a crossing critical point.
 The estimate of $T_c=0.603(2)$ and
$\nu = 1.50(1)$ is  consistent with that obtained 
 earlier.

%--------------------------------Fig11--------------------------------------
%------------------------------------------------------------------------------

%--------------------------------Fig12--------------------------------------

%------------------------------------------------------------------------------

After obtaining the critical exponents, we can now discuss 
the universality classes of the phase transitions.
Two symmetry breakings are obvious from
 the snapshot series of magnetization shown in Fig. \ref{SNAP}
as well as from temperature dependence of $M_i$ in Fig. \ref{Ni}. 
At higher temperature, the native octahedral symmetry
$O_h$ breaks into an intermediate phase  $C_{3h}$ symmetry which
then freezes into a ground state of low temperatures.
 The high temperature phase transition with exponents
 $\nu_h = 1.50(1)$ and $\eta_h = 0.260(1)$ 
different from Ising's exponents may suggest the 
existence of cubic universality class in two dimensions. 

We expect the low temperature transition is 
in the same universality class of 3-state Potts model,
which is a realization of $C_{3h}$ symmetry and exactly
solvable with exponents $\nu = 5/6$ and $\eta =
 4/15$ \cite{wu}. 
Our numerical results affirm this scenario.

\section{Concluding Remarks}\label{four}

In summary, we have investigated 
the ferromagnetic  edge-cubic spin model
on square lattice with periodic boundary
condition. The octahedral symmetry group
$O_h$ of the system experiences sequential
symmetry breakings as temperature is reduced.
Firstly, the $O_h$ breaks into
$C_{3h}$ which occurs at critical temperature $0.602(1)$.
Two critical exponents are estimated, i.e., the
exponent of the correlation length  $\nu=1.50(1)$
and decay exponent of correlation function $\eta=0.260(1).$

Further cooling down the system, the second
phase transition is observed.
Although  the magnetization is the order 
parameter for the ferromagnetic system,
it does not necessarily succeed in the analysis of 
low temperature transition of our system. 
The introduced order
parameter associated with the maximum number of spins 
pointing to a particular direction, in fact performs better,
by which we extract the critical temperature and exponents.

The low temperature  transition that occurs at $ T = 0.5422(2)$ 
separates between the intermediate state belonging to symmetry
group $C_{3h}$ and the ground state. Two
critical exponents of this transition are estimated, namely
$\nu$ of correlation length and $\eta$ of decaying correlation
function, tabulated in Table \ref{table1}.  
The values of the exponents
are in very good agreement with the 3-state Potts model.

\begin{table}
\caption{Transition temperatures and exponents $\nu$
and $\eta$ of high and low temperature transition.$\\$}
\label{table1}
\begin{tabular}{c|c|c|c}
\hline
\hline
Transition & $T_c$ & $\nu$& $\eta$\\
\hline
High-$T$ & $0.602(1)$& 1.50(1) & 0.260(1)\\
Low-$T$ & $0.5422(2)$& 0.833(1) &  0.267(1)\\
\hline
\end{tabular}
\end{table}

\section*{Acknowledgments}

The authors wish to thank  Y. Tomita and T. Suzuki for valuable discussions.  
They also thank D. Ueno for the collaboration in the early
stage of research.  The extensive computation was
performed using the  supercomputer facilities of
 the Institute of Solid State Physics, University of Tokyo, Japan.
The present work is financially supported by KAKENHI 19340109
and 19052004, and by Next Generation Supercomputing Project, 
Nanoscience Program, MEXT, Japan.

\end{document}